\documentclass[prl,aps,amsmath,amssymb,amsfonts,showpacs,superscriptaddress,twocolumn]{revtex4}
\input{epsf}
\usepackage{psfrag,graphicx}
\usepackage{epsfig}
\def\tstar{T_F^\star}
\def\Jaf{J_{\rm{AF}}\,}
\def\smax{$s_{\rm{max}}$}

\begin{document}

\title{Interaction-induced adiabatic cooling and antiferromagnetism
of cold fermions in optical lattices}
\author{F. Werner}
\affiliation{LKB, Ecole Normale Sup\'erieure, 24 Rue Lhomond, 75231 Paris Cedex 05,
France} \affiliation{CPHT, Ecole Polytechnique, 91128 Palaiseau
Cedex, France}
\author{O. Parcollet}
\affiliation{SPhT, CEA-Saclay, 91191 Gif sur Yvette Cedex}
\author{A. Georges}
\affiliation{CPHT, Ecole Polytechnique, 91128 Palaiseau Cedex,
France}
\author{S. R. Hassan}
\affiliation{CPHT, Ecole Polytechnique, 91128 Palaiseau Cedex,
France}
\begin{abstract}
We propose an interaction-induced cooling mechanism for two-component cold fermions
in an optical lattice. It is based on an increase of the ``spin'' entropy upon localisation,
an analogue of the Pomeranchuk effect in liquid Helium 3. We discuss its application
to the experimental realisation of the antiferromagnetic phase. We illustrate our arguments
with Dynamical Mean-Field Theory calculations.
\end{abstract}

 \maketitle

Cold atoms in optical lattices~\cite{jaksch_toolbox} offer a promising laboratory for the study
of strongly correlated systems, bringing quantum optics to have bearing on
key issues in condensed matter physics.
Pioneering experiments on the Mott insulator to superfluid
transition~\cite{greiner_mott_nature_2002}
have demonstrated the possibility~\cite{jaksch_lattice_prl_1998} of probing quantum phase transitions
between different ground-states of these systems. Recently, great progress
has been achieved on cold Fermi gases as well, resulting in the production
of molecular condensates
in trapped gases~\cite{greiner_bec_mol_nature_2003,jochim_bec_mol_science_2003,
zwierlein_bec_mol_prl_2003,bourdel_bec_bcs_prl_2004} and the first imaging of Fermi surfaces in a
three-dimensional optical lattice~\cite{kohl_fermisurface_prl_2005}. Controllability
is one of the most remarkable aspects of these systems, with the possibility of
tuning both the tunneling amplitude between lattice sites ($t$) and the on-site interaction
strength ($U$), by varying the depth of the
optical lattice, and by varying the inter-atomic scattering length thanks to
Feschbach resonances.

In this letter, we consider fermionic atoms with two hyperfine (``spin'') states in an
optical lattice. When the lattice is deep and the scattering length is small (see below for
a precise
condition), a one-band Hubbard model is realized.
The main physical effect studied in this paper is
the possibility of cooling down the system by increasing the interaction
strength adiabatically.
As described below, this is due to a higher degree of localization -and hence
an increase in spin entropy- as $U/t$ or the temperature is increased.
This is a direct analogue of the
Pomeranchuk effect in liquid Helium 3. This mechanism relies on interactions and
should be distinguished from the adiabatic cooling for non-interacting atoms in the lattice
discussed in ~\cite{blakie_cooling_fermions,Demler_Cirac_Zoller}.
The second main goal of the present paper is to study how this
effect can be used in order to reach the phase with
antiferromagnetic long-range order. For deep lattices (large
$U/t$), the N\'eel temperature is expected to become very low, of
the order of the magnetic superexchange $J_{\rm{AF}}= 4t^2/U$.
Naively, it would seem that this requires extreme cooling of the
gas. Here, we point out that the appropriate concept is actually
the entropy along the antiferromagnetic critical line, and that at
large $U/t$ this quantity tends to {\it a finite constant} which
depends only on the specific lattice. Hence, cooling the gas down
to a temperature corresponding to this finite entropy per atom,
and then following equal-entropy trajectories, should be enough
to reach the magnetic phase.
These physical observations are substantiated by theoretical calculations
using in particular dynamical mean-field theory
(DMFT)~\cite{georges_review_dmft,georges_strong},
an approach that has led to important progress on strongly correlated
fermion systems in recent years.

\begin{figure}
\includegraphics[width=\columnwidth,clip=]{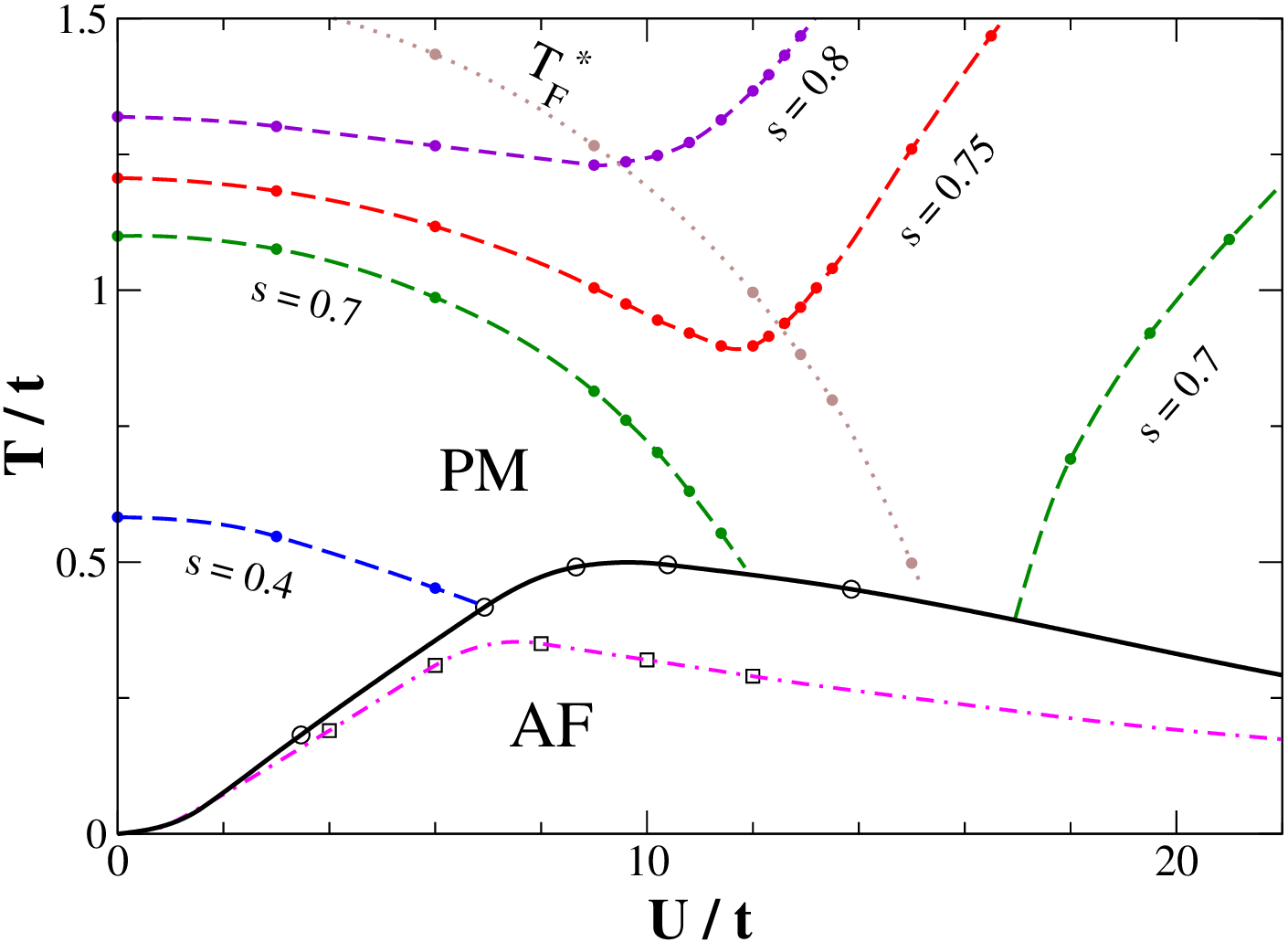}
\caption{(Color online) Phase diagram of the half-filled Hubbard model on the cubic lattice:
antiferromagnetic (AF) and paramagnetic (PM) phase. Transition temperature
within DMFT approximation (plain curve, open circles) and QMC calculation of
Ref.~\cite{staudt_af_hubbard_epjb_2000} (dot-dashed curve, squares).
Dashed lines: isentropic curves (s=0.4,0.7,0.75,0.8), computed within DMFT.
Dotted line: quasiparticle coherence scale $T_F^*(U)$.
The DMFT results were obtained with QMC (for $T_N$) and the IPT
approximation\cite{georges_review_dmft} (for the isentropics). The transition
curves are interpolations, continued at high $U/t$ using the analytical
expressions for the Heisenberg
regime.
}
\label{fig:T_N and isentropics}
\end{figure}

We consider the one-band repulsive Hubbard model:
\begin{equation}
H = - \sum_{i,j,\sigma} t_{ij}\,c^\dag_{i\sigma} c_{j\sigma}
+ U \sum_i n_{i\uparrow} n_{i\downarrow}
\label{Hubbard H}
\end{equation}
where $i,j$ are site indices on the lattice, and $\sigma = \uparrow,\downarrow$ is a
``spin'' index associated with the two hyperfine states.
The conditions under which
two-component fermionic atoms in an optical lattice actually realize such a
single-band lattice model will be discussed later.
On an unfrustrated bipartite
three-dimensional lattice (e.g. the cubic lattice),
with hopping between nearest-neighbor sites $t_{ij}=t$, and
for one particle per site on average (half-filling), the physics of this model is
rather well understood (see e.g.~\cite{staudt_af_hubbard_epjb_2000}).
For temperatures above the N\'eel critical temperature $T_N$,
the system is a paramagnet with an increasing tendency to Mott localization as
$U/t$ is increased (the Mott gap becomes of order $U$ at large $U/t$).
For $T<T_N$, the antiferromagnetic phase (Fig.~\ref{fig:T_N and isentropics}) displays two-sublattice
spin ordering and a doubling of the unit-cell. At
weak coupling (small $U/t$), this is a spin-density wave instability with a weak modulation of
the
sublattice magnetization. In this regime, $T_N$ is exponentially small in $t/U$, as a simple
Hartree mean-field theory suggests. At strong coupling (large $U/t$), the low-energy sector of the
model is described by a Heisenberg exchange Hamiltonian $\Jaf\sum_{<ij>} \vec{S}_i\cdot\vec{S}_j$
with $\Jaf=4t^2/U$. In this Heisenberg limit, $T_N=\theta\Jaf$, with $\theta$ a numerical constant depending on the
lattice ($\theta=0.957$ for the cubic lattice \cite{staudt_af_hubbard_epjb_2000}).
These two regimes are connected by a smooth crossover
(which is equivalent to the BEC-BCS crossover at half filling).
The N\'eel temperature displays a maximum at intermediate coupling, as a function of $U/t$.
This is illustrated by Fig.~\ref{fig:T_N and isentropics}, on which we display our calculation of
$T_N$ vs. $U/t$, using the DMFT approximation on the cubic lattice and the quantum
Monte-Carlo (QMC) Hirsch-Fye algorithm.
DMFT overestimates
$T_N$ by about $50\%$ in the intermediate coupling regime,
in comparison to the direct QMC calculations
of Ref.~\cite{staudt_af_hubbard_epjb_2000} on the cubic lattice (also displayed on
Fig.~\ref{fig:T_N and isentropics}).

\begin{figure}
\includegraphics[width=8 cm,clip=]{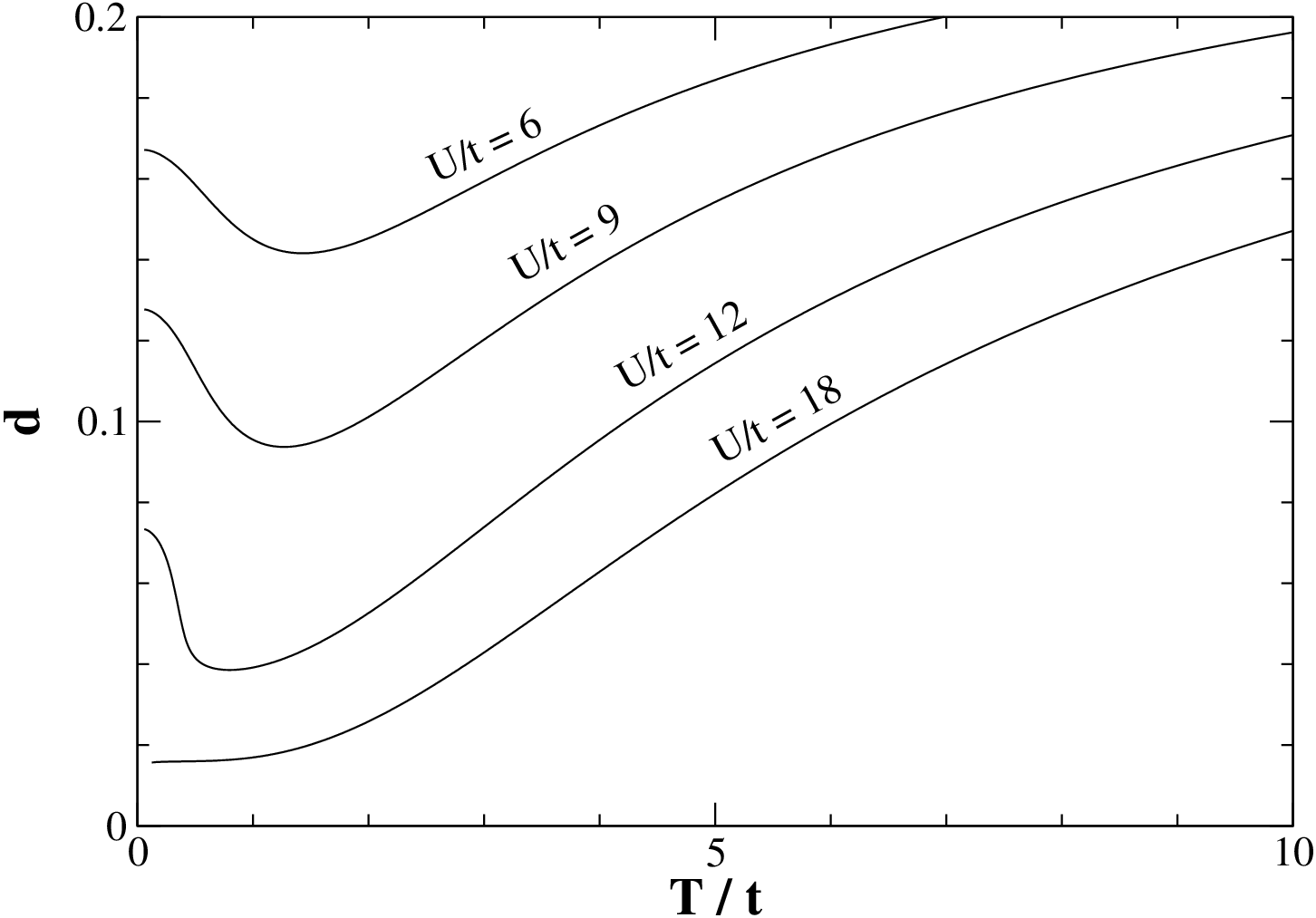}
\caption{Double occupancy $d=\left< n_{i\uparrow} n_{i\downarrow} \right>$
as a function of temperature, for several values of $U/t$, calculated within
DMFT(IPT).
The initial decrease is the Pomeranchuk effect responsible for adiabatic cooling.}
\label{fig:double_occupancy}
\end{figure}

We now discuss how the entropy varies as the effective strength of the
on-site interaction $U/t$ is changed in the paramagnetic phase.
Since all properties
depend on the ratios $T/t$ and $U/t$, we can consider that the hopping is
fixed and that $T$ and $U$ are varied, or alternatively that both the temperature and
coupling are measured in units of $t$, the natural unit of kinetic energy.
Denoting by $f$ and $s$ the free-energy and entropy
per lattice site, respectively, one has: $s= -\partial f/\partial T$ and $\partial f/\partial
U=d$, with $d$ the probability that a given site is doubly occupied:
$d\equiv\langle n_{i\uparrow} n_{i\downarrow} \rangle$. We thus obtain:
\begin{equation}
\frac{\partial s}{\partial U} = - \frac{\partial d}{\partial T}
\label{eq:partial s}
\end{equation}
This equation can be used to discuss qualitatively the shape of the isentropic
curves $T_i=T_i(U)$ in the $(U,T)$ phase diagram, along which $s(T_i(U),U)=\rm{const.}$
Taking a derivative of this equation yields:
\begin{equation}
c(T_i)\,\frac{\partial T_i}{\partial U} =
T_i\,\frac{\partial d}{\partial T}|_{T=T_i}
\label{eq:isentropics}
\end{equation}
in which $c=T\partial s/\partial T$ is the specific heat per lattice
site.  Fortunately, the temperature-dependence of the probability of
double occupancy $d(T)$ has been studied in previous work by one of
the authors~\cite{georges_krauth_mott_prl,georges_krauth_mott_prb} and
others~\cite{rozenberg_finiteT_mott}.
It was observed that, when $U/t$ is not too large, the double
occupancy first {\it decreases} as temperature is increased from $T=0$
(indicating a higher degree of localisation), and
then turns around and grows again.
This is shown on Fig.~\ref{fig:double_occupancy} using DMFT calculations.
This apparently counter-intuitive behavior is a direct
analogue of the Pomeranchuk effect in liquid Helium 3: since the
(spin-) entropy is larger in a localised state than when the fermions
form a Fermi-liquid (in which $s\propto T$), it is favorable to
increase the degree of localisation upon heating. The minimum of
$d(T)$ essentially coincides with the quasiparticle coherence scale
$\tstar(U)$ which is a rapidly decreasing function of $U$
(Fig.~\ref{fig:T_N and isentropics}).  This phenomenon therefore
applies only as long as $\tstar>T_N$, and hence when $U/t$ is not too
large. For large $U/t$, Mott localisation dominates for all
temperatures $T<U$ and suppresses this effect.
Since $\partial d/\partial T<0$ for $T<\tstar(U)$ while $\partial
d/\partial T>0$ for $T>\tstar(U)$, Eq.(\ref{eq:isentropics})
implies that the isentropic curves of the half-filled Hubbard
model (for not too high values of the entropy) must have a
negative slope at weak to
intermediate coupling, before turning around at stronger coupling.
In order to substantiate this behavior, inferred on rather general
grounds, we have performed DMFT calculations of the isentropic curves,
with results displayed in Fig.\ref{fig:T_N and isentropics}.
The entropy $s(T)$ was calculated by integrating the internal energy per site $e(T)$
according to: $s(T)=\ln 4+e(T)/T-\int_T^\infty dT' e(T')/T'^2$, which
follows from the thermodynamic relation $\partial_T e=T\partial_T s$.
The DMFT equations were solved using the ``iterated perturbation theory'' (IPT)
approximation\cite{georges_review_dmft} (using, for simplicity, a semicircular density of states), and the
internal energy was calculated
from the one-particle Green's function.

It is clear from the results of Fig.~\ref{fig:T_N and isentropics}
that, starting from a low enough initial value of the entropy per site,
adiabatic cooling can be achieved by
either increasing $U/t$ starting from a small value, or decreasing $U/t$ starting from
a large value (the latter requires however to cool down the gas
while the lattice is already present).
We emphasize that this cooling mechanism is an interaction-driven
phenomenon: indeed, as $U/t$ is increased, it allows to lower the
{\it reduced temperature} $T/t$, normalized to the natural scale
for the Fermi energy in the presence of the lattice. Hence, cooling is not
simply due to the tunneling amplitude $t$ becoming
smaller as the lattice is turned on.
At weak coupling and low temperature,
the cooling mechanism can be related to the effective mass of
quasiparticles ($\propto 1/\tstar$) becoming heavier as $U/t$ is increased,
due to Mott localisation. Indeed, in this regime,
the entropy is proportional to $T/\tstar(U)$. Hence,
conserving the entropy while increasing $U/t$ adiabatically from
$(U/t)_i$ to $(U/t)_f$ will reduce the final temperature in comparison to
the initial one $T_i$ according to: $T_f/T_i=\tstar(U_f)/\tstar(U_i)$.

At this stage, let us briefly discuss the validity of the DMFT approach,
extensively used in the present work. In this approach, the lattice model is
mapped onto a single-site quantum problem coupled to a self-consistent
effective medium. This is an approximation, which becomes exact only in the
limit of infinite lattice coordination~\cite{georges_review_dmft}.
As a local approach, it underestimates the precursor antiferromagnetic
correlations above $T_N$, which will in turn quench the entropy and
ultimately play against the cooling mechanism very close to $T_N$.
However, as long as the correlation length is not too large,
a local approximation should be accurate. Indeed, the existence of
a minimum in $d(T)$ has been confirmed by the calculations of Ref.~\cite{DareAlbinet}
using a different method, for a three-dimensional lattice, suggesting that
the cooling mechanism discussed here is a robust effect.

\begin{figure}
\includegraphics[width=8 cm,clip=]{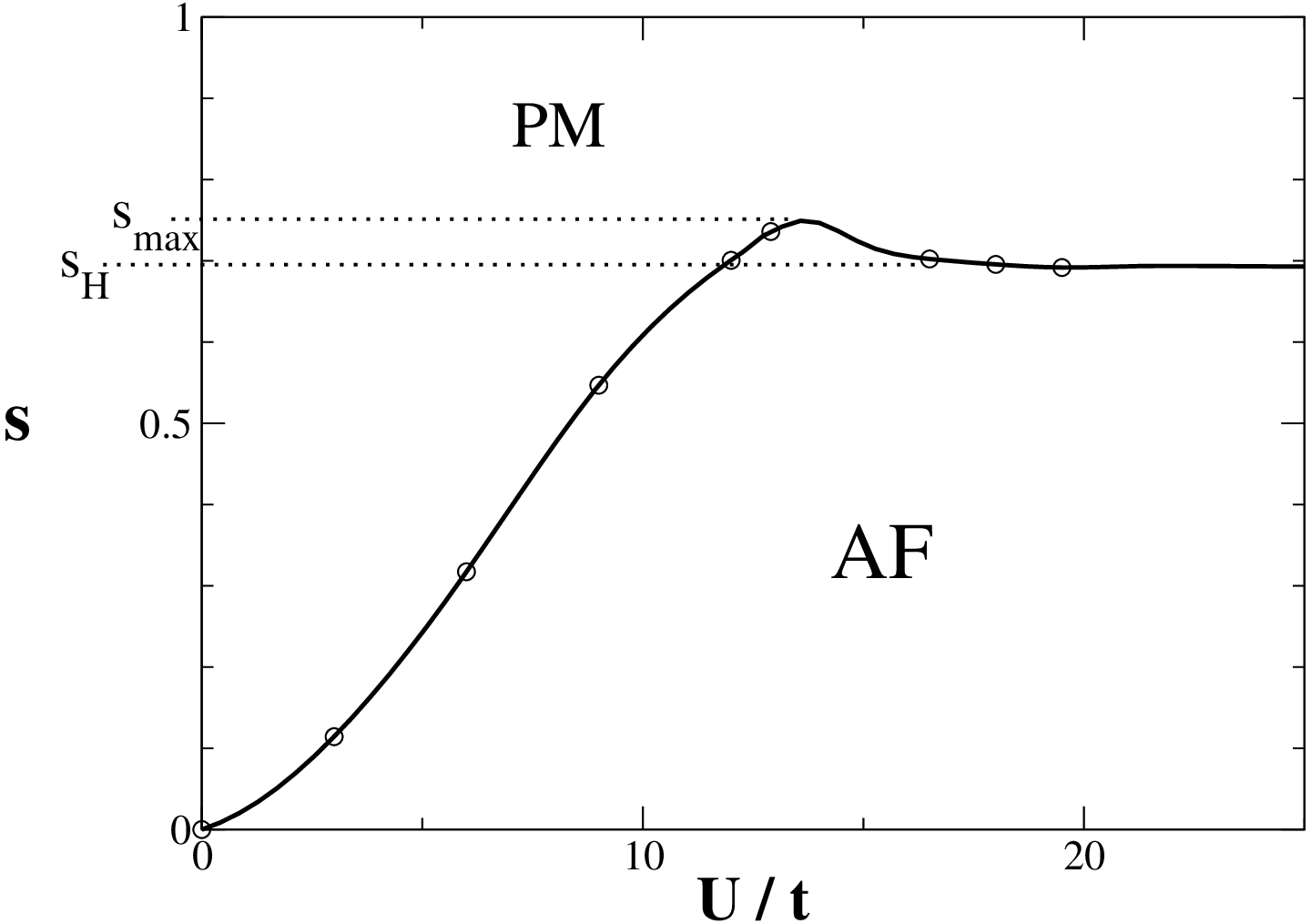}
\caption{Phase diagram as a function of entropy. The displayed curve results
from a DMFT-IPT calculation (in which case
$s_H=\ln 2$), but its shape is expected to be general
(with $s_H$ reduced by quantum fluctuations).}
\label{fig:s_N}
\end{figure}

The isentropic curves in Fig.~\ref{fig:T_N and isentropics} suggest
that interaction-induced adiabatic cooling could be used in order to
reach the magnetically ordered phase.
To explore this idea in more details, we focus on the entropy along the N\'eel critical
boundary $s_N(U)\equiv s(T_N(U),U)$.
At weak-coupling (spin-density wave regime),
$s_N(U)$ is expected to be exponentially small. In contrast, in the opposite Heisenberg regime of
large $U/t$, $s_N$ will reach a finite value $s_H$, which is the entropy of the quantum Heisenberg
model at its critical point. $s_H$ is a pure number which depends only on the specific lattice
of interest. Mean-field theory of the Heisenberg model yields $s_H=\ln 2$, but quantum
fluctuations will reduce this number. We have performed a Schwinger boson calculation of
this quantity, along the lines of \cite{ArovasAuerbach,BOOK_Auerbach}, and found that
this reduction is of the order of $50\%$ on the cubic lattice.
How does $s_N$ evolve from weak to strong coupling~? A rather general argument suggests that
it should go through a maximum \smax$>s_H$. In order to see this, we use again
(\ref{eq:partial s})
and take a derivative of $s_N=s(T_N(U),U))$,
which yields:
\begin{equation}
\frac{ds_N}{dU} =
\frac{c(T_N)}{T_N}\,\frac{dT_N}{dU} - \frac{\partial d}{\partial T}|_{T=T_N}
\label{eq:delsU}
\end{equation}
If only the first term was present in the r.h.s of this equation, it would imply
that $s_N$ is maximum exactly at the value of the coupling where $T_N$ is
maximum (note that
$c(T_N)$ is finite ($\alpha<0$)
for the 3D-Heisenberg model \cite{ChenFerrenbergLandau}).
However, in view of the above properties of the double occupancy,
the second term in the r.h.s has a similar variation than the first one: it starts
positive, and then changes sign at an intermediate coupling when $\tstar(U)=T_N(U)$. These considerations
suggest that $s_N(U)$ does reach a maximum
value at intermediate coupling, in the same regime where $T_N$ reaches a maximum.
Hence, $s_N(U)$ has the general form sketched on Fig.~\ref{fig:s_N}.
This figure can be viewed as a phase diagram of the half-filled Hubbard model, in
which {\it entropy itself is used as a thermometer}, a very natural
representation when addressing adiabatic cooling.
Experimentally, one may first cool down the gas (in the absence of the optical
lattice) down to a temperature where the entropy per particle is lower than
$s_H$ (this corresponds to $T/T_F<s_H/\pi^2$ for a trapped ideal gas).
Then, by branching on the optical lattice adiabatically, one could increase
$U/t$ until one particle per site is reached over most of the trap: this should allow
to reach the antiferromagnetic phase. Assuming that the timescale for
adiabaticity is simply set by the hopping, we observe that typically
$\hbar/t\sim 1 \textrm{ms}$.

\begin{figure}
\includegraphics[width=\columnwidth,clip=]{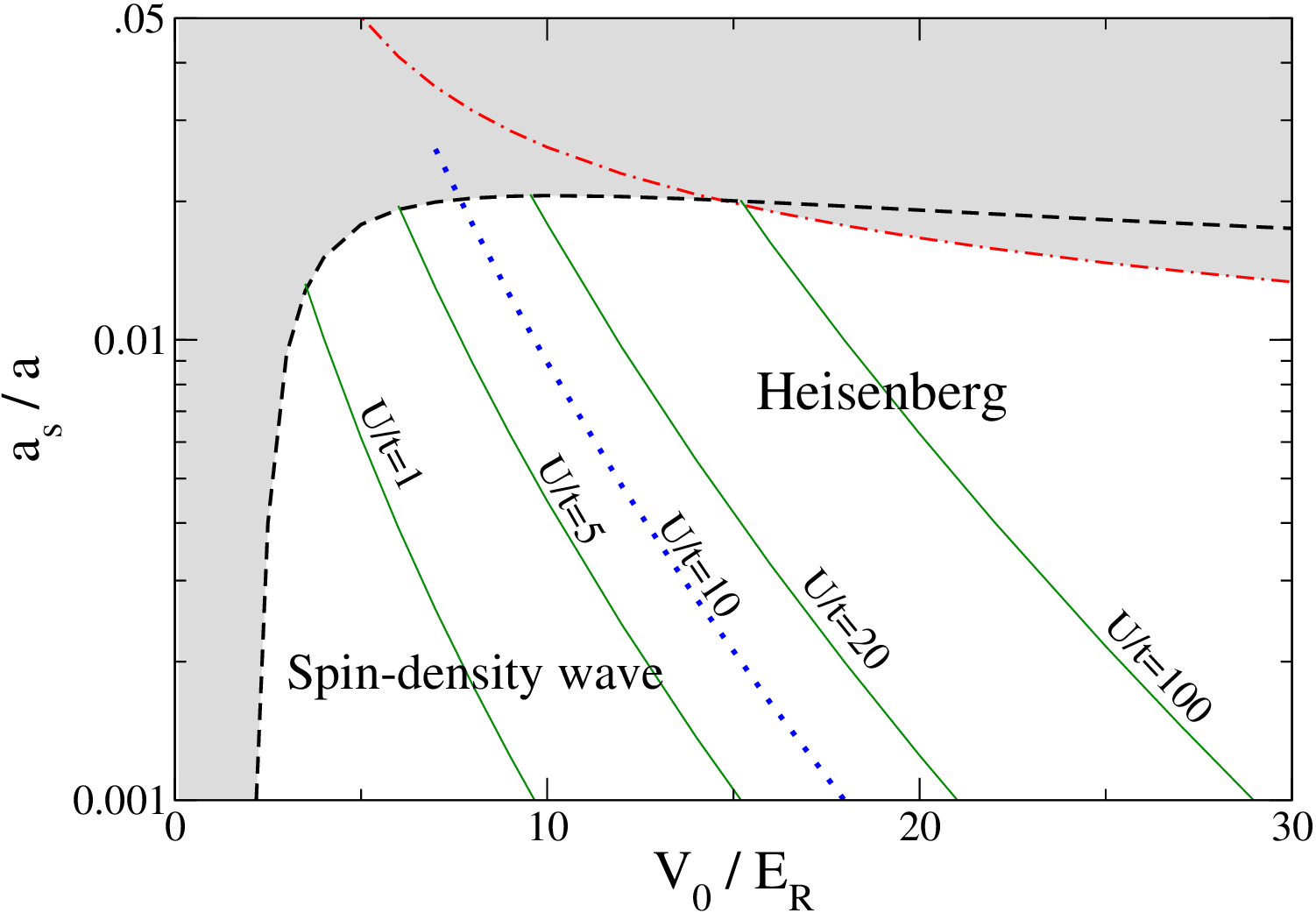}
\caption{Spin-density wave and Heisenberg regimes as a function of the
depth of the periodic potential $V_0$
and the scattering length $a_s$.
The crossover between these regimes is indicated by the dotted line ($U/t=10$),
where $T_N/t$ is maximum (other contour lines are also indicated).
In the shaded region, the one-band Hubbard description is no longer valid.
Above the dashed line ($U/\Delta > 0.1$) other bands must be taken into account and the
pseudopotential approximation fails. Above the dashed-dotted line,
non-Hubbard interaction terms become sizeable ($t_d/t > 0.1$, see text).
}
\label{limites_validite}
\end{figure}
Let us now discuss the conditions under which two-component fermions
in an optical lattice are accurately described by the Hubbard hamiltonian
(\ref{Hubbard H}) (see also \cite{jaksch_toolbox,jaksch_lattice_prl_1998}).
The many-body hamiltonian is written in second-quantized form using
as single-particle basis functions the Wannier functions associated
with the periodic potential
$V_{opt}(\vec{r}) = V_0 \sum_{i=1}^3 \sin^2(\pi x_i /a)$
(the lattice spacing is $a=\lambda/2$, with $\lambda$ the wavelength of the
laser).
The interaction terms are obtained as matrix elements of the low-energy effective
potential
$V_{int}(\vec{r_1}-\vec{r_2})=\frac{4\pi\hbar^2 a_s}{m} \,
\delta^3(\vec{r_1}-\vec{r_2})$ where $a_s$ is the scattering length.
In general, this results in a multi-band model which, besides the on-site Hubbard interaction,
involves also more complicated interaction
terms such as nearest-neighbor interactions, or
density-assisted hopping terms of the form $t_d c^\dagger_i c_j n_i$,
with $i,j$ neighboring sites.
By explicitly computing these terms, as well as the one-body part of the
hamiltonian, we examined under which conditions (i) the reduction to a one-band
model is valid and (ii) these non-Hubbard interactions are negligible.
This determines a domain in the $(V_0/E_R,a_s/a)$ plane (with
$E_R=\hbar^2 \pi^2 /2ma^2$ the recoil energy), which is depicted
on Fig.~\ref{limites_validite}.
Condition (i) requires in particular that
the on-site Hubbard repulsion is smaller than the gap $\Delta$
between the first and the second band: $U \ll \Delta$. At large values
of $V_0/E_R$, it can be shown that this is also the condition for our use of the pseudopotential
approximation to be valid:
$a_s\ll l_0$ with $l_0$ the spatial
extension of the Wannier function of the first band.
We found that the stricter condition of type (ii) originates from density-assisted
hopping terms which should obey $t_d\ll t$.
We also displayed on Fig.~\ref{limites_validite} some contour lines associated with a
given value of $U/t$. The one associated with $U/t\approx 10$ can be taken as
the approximate separatrix between the spin density-wave
and Heisenberg antiferromagnetic regions.
$T_N/t$ is maximal along this line, and $T_N<0.015 E_R$ in the allowed
region. Thus adiabatic cooling is important to reach the AF phase.
Since $V_0$ and $a_s$ are the two experimentally tunable parameters,
Fig.~\ref{limites_validite} aims at summarizing useful information for such
experimental investigations.
The detection of the antiferromagnetic long-range order might be achieved by
spin-selective Bragg spectroscopy in order to reveal the doubling of the unit cell.
The two hyperfine states could be distinguished by their Zeeman splitting,
or using polarised light.
A different method, which has been recently proposed~\cite{altman_2004} and investigated
experimentally~\cite{folling}, is to use quantum noise
interferometry.

To summarize, we propose in this article  an interaction-induced cooling mechanism for
two-component cold fermions in an optical lattice. One possible application of this
mechanism is in reaching the phase with antiferromagnetic long-range order.

\acknowledgments
We are grateful to C.~Salomon and F.~Chevy for encouragement and several
discussions. We also thank S. Biermann, I.~Bloch, Y. Castin, J. Dalibard, E. Demler, W. Krauth,
F. Lechermann, L. Tarruell and A.-M. Tremblay. We acknowledge the support of an
AC-Nanosciences
``Gaz quantiques'' (project number 201), the IFCPAR (project number 2404-1), CNRS and Ecole Polytechnique.


\begin{thebibliography}{20}
\expandafter\ifx\csname
natexlab\endcsname\relax\def\natexlab#1{#1}\fi
\expandafter\ifx\csname bibnamefont\endcsname\relax
  \def\bibnamefont#1{#1}\fi
\expandafter\ifx\csname bibfnamefont\endcsname\relax
  \def\bibfnamefont#1{#1}\fi
\expandafter\ifx\csname citenamefont\endcsname\relax
  \def\citenamefont#1{#1}\fi
\expandafter\ifx\csname url\endcsname\relax
  \def\url#1{\texttt{#1}}\fi
\expandafter\ifx\csname
urlprefix\endcsname\relax\def\urlprefix{URL }\fi
\providecommand{\bibinfo}[2]{#2}
\providecommand{\eprint}[2][]{\url{#2}}

\bibitem[{\citenamefont{{Jaksch} and {Zoller}}(2004)}]{jaksch_toolbox}
\bibinfo{author}{\bibfnamefont{D.}~\bibnamefont{{Jaksch}}} \bibnamefont{and}
  \bibinfo{author}{\bibfnamefont{P.}~\bibnamefont{{Zoller}}}
  (\bibinfo{year}{2004}),
  \eprint{cond-mat/0410614}.

\bibitem[{\citenamefont{Greiner et~al.}(2002)\citenamefont{Greiner, Mandel,
  Esslinger, H\"ansch, and Bloch}}]{greiner_mott_nature_2002}
 M.~Greiner et al.,
  \bibinfo{journal}{Nature} \textbf{\bibinfo{volume}{415}}, \bibinfo{pages}{39}
  (\bibinfo{year}{2002}).

\bibitem[{\citenamefont{{Jaksch} et~al.}(1998)\citenamefont{{Jaksch}, {Bruder},
{Cirac}, {Gardiner}, and {Zoller}}}]{jaksch_lattice_prl_1998}
D.~Jaksch et al.,
  \bibinfo{journal}{Phys. Rev. Lett.} \textbf{\bibinfo{volume}{81}},
  \bibinfo{pages}{3108} (\bibinfo{year}{1998}).

\bibitem[{\citenamefont{Greiner et~al.}(2003)\citenamefont{Greiner, Regal, and
  Jin}}]{greiner_bec_mol_nature_2003}
M.~Greiner et al.,
  \bibinfo{journal}{Nature} \textbf{\bibinfo{volume}{537}},
  \bibinfo{pages}{426} (\bibinfo{year}{2003}).

\bibitem[{\citenamefont{{Jochim} et~al.}(2003)\citenamefont{{Jochim},
  {Bartenstein}, {Altmeyer}, {Hendl}, {Riedl}, {Chin}, {Denschlag}, and
  {Grimm}}}]{jochim_bec_mol_science_2003}
  S.~Jochim et~al.,
  \bibinfo{journal}{Science} \textbf{\bibinfo{volume}{302}},
  \bibinfo{pages}{2101} (\bibinfo{year}{2003}).

\bibitem[{\citenamefont{{Zwierlein} et~al.}(2003)\citenamefont{{Zwierlein},
  {Stan}, {Schunck}, {Raupach}, {Gupta}, {Hadzibabic}, and
  {Ketterle}}}]{zwierlein_bec_mol_prl_2003}
M.~W.~Zwierlein et al.,
  \bibinfo{journal}{Phys. Rev. Lett.}
  \textbf{\bibinfo{volume}{91}}(\bibinfo{number}{25}), \bibinfo{pages}{250401}
  (\bibinfo{year}{2003}).

\bibitem[{\citenamefont{{Bourdel} et~al.}(2004)\citenamefont{{Bourdel},
  {Khaykovich}, {Cubizolles}, {Zhang}, {Chevy}, {Teichmann}, {Tarruell},
  {Kokkelmans}, and {Salomon}}}]{bourdel_bec_bcs_prl_2004}
T.~Bourdel et al.,
  \bibinfo{journal}{Phys. Rev. Lett.}
  \textbf{\bibinfo{volume}{93}}(\bibinfo{number}{5}), \bibinfo{pages}{050401}
  (\bibinfo{year}{2004}).

\bibitem[{\citenamefont{{K{\" o}hl} et~al.}(2004)\citenamefont{{K{\" o}hl},
  {Moritz}, {St{\" o}ferle}, {G{\" u}nter}, and
  {Esslinger}}}]{kohl_fermisurface_prl_2005}
M.~K\"{o}hl et al.,
  (\bibinfo{year}{2004}),
  \eprint{cond-mat/0410389}.

\bibitem[{\citenamefont{{Blakie} and {Bezett}}(2004)}]{blakie_cooling_fermions}
\bibinfo{author}{\bibfnamefont{P.~B.} \bibnamefont{{Blakie}}} \bibnamefont{and}
  \bibinfo{author}{\bibfnamefont{A.}~\bibnamefont{{Bezett}}} (\bibinfo{year}{2004}),
  \eprint{cond-mat/0410140}.

\bibitem[{\citenamefont{Hofstetter et~al.}(2002)\citenamefont{Hofstetter,
  Cirac, Zoller, Demler, and Lukin}}]{Demler_Cirac_Zoller}
W.~Hofstetter et al.,
  \bibinfo{journal}{Phys. Rev. Lett.} \textbf{\bibinfo{volume}{89}},
  \bibinfo{pages}{220407} (\bibinfo{year}{2002}).

\bibitem[{\citenamefont{{Georges} et~al.}(1996)\citenamefont{{Georges},
  {Kotliar}, {Krauth}, and {Rozenberg}}}]{georges_review_dmft}
A.~Georges et al.,
  \bibinfo{journal}{Rev. Mod. Phys.} \textbf{\bibinfo{volume}{68}},
  \bibinfo{pages}{13} (\bibinfo{year}{1996}).

\bibitem[{\citenamefont{{Georges}}(2004)}]{georges_strong}
\bibinfo{author}{\bibfnamefont{A.}~\bibnamefont{{Georges}}}, in
  \emph{\bibinfo{booktitle}{Lectures on the physics of highly correlated
  electron systems VIII}}, edited by
  \bibinfo{editor}{\bibfnamefont{A.}~\bibnamefont{Avella}} \bibnamefont{and}
  \bibinfo{editor}{\bibfnamefont{F.}~\bibnamefont{Mancini}}
  (\bibinfo{publisher}{American Institute of Physics}, \bibinfo{year}{2004}),
  \bibinfo{note}{cond-mat/0403123}.

\bibitem[{\citenamefont{Staudt et~al.}(2000)\citenamefont{Staudt, Dzierzawa,
  and Muramatsu}}]{staudt_af_hubbard_epjb_2000}
R.~Staudt et al.,
  \bibinfo{journal}{Eur. Phys. J. B} \textbf{\bibinfo{volume}{17}},
  \bibinfo{pages}{411} (\bibinfo{year}{2000}).

\bibitem[{\citenamefont{{Georges} and
  {Krauth}}(1992)}]{georges_krauth_mott_prl}
\bibinfo{author}{\bibfnamefont{A.}~\bibnamefont{{Georges}}} \bibnamefont{and}
  \bibinfo{author}{\bibfnamefont{W.}~\bibnamefont{{Krauth}}},
  \bibinfo{journal}{Phys. Rev. Lett.} \textbf{\bibinfo{volume}{69}},
  \bibinfo{pages}{1240} (\bibinfo{year}{1992}).

\bibitem[{\citenamefont{{Georges} and
  {Krauth}}(1993)}]{georges_krauth_mott_prb}
\bibinfo{author}{\bibfnamefont{A.}~\bibnamefont{{Georges}}} \bibnamefont{and}
  \bibinfo{author}{\bibfnamefont{W.}~\bibnamefont{{Krauth}}},
  \bibinfo{journal}{Phys. Rev. B} \textbf{\bibinfo{volume}{48}},
  \bibinfo{pages}{7167} (\bibinfo{year}{1993}).

\bibitem[{\citenamefont{Rozenberg et~al.}(1999)\citenamefont{Rozenberg, Chitra,
  and Kotliar}}]{rozenberg_finiteT_mott}
M.~J.~Rozenberg et al.,
  \bibinfo{journal}{Phys. Rev. Lett.} \textbf{\bibinfo{volume}{83}},
  \bibinfo{pages}{3498} (\bibinfo{year}{1999}).

\bibitem[{\citenamefont{Dar\'e and Albinet}(2000)}]{DareAlbinet}
\bibinfo{author}{\bibfnamefont{A.-M.} \bibnamefont{Dar\'e}} \bibnamefont{and}
  \bibinfo{author}{\bibfnamefont{G.}~\bibnamefont{Albinet}},
  \bibinfo{journal}{Phys. Rev. B.} \textbf{\bibinfo{volume}{61}},
  \bibinfo{pages}{4567} (\bibinfo{year}{2000}).

\bibitem[{\citenamefont{Arovas and Auerbach}(1988)}]{ArovasAuerbach}
\bibinfo{author}{\bibfnamefont{D.}~\bibnamefont{Arovas}} \bibnamefont{and}
  \bibinfo{author}{\bibfnamefont{A.}~\bibnamefont{Auerbach}},
  \bibinfo{journal}{Phys. Rev. B.} \textbf{\bibinfo{volume}{38}},
  \bibinfo{pages}{316} (\bibinfo{year}{1988}).

\bibitem[{\citenamefont{Auerbach}(1994)}]{BOOK_Auerbach}
\bibinfo{author}{\bibfnamefont{A.}~\bibnamefont{Auerbach}},
  \emph{\bibinfo{title}{Interacting Electrons and Quantum Magnetism}}
  (\bibinfo{publisher}{Springer}, \bibinfo{year}{1994}).

\bibitem[{\citenamefont{Chen et~al.}(1993)\citenamefont{Chen, Ferrenberg, and
  Landau}}]{ChenFerrenbergLandau}
K.~Chen et al.,
  \bibinfo{journal}{Phys. Rev. B.} \textbf{\bibinfo{volume}{48}},
  \bibinfo{pages}{3249} (\bibinfo{year}{1993}).

\bibitem{altman_2004} E. Altman et al., Phys. Rev. A \textbf{70},
013603 (2004).

\bibitem{folling} S.~F\"{o}lling et~al., Nature \textbf{434}, 481 (2005).

\end{thebibliography}
\end{document}